\documentclass[twocolumn,showpacs,showkeywords,preprintnumbers,amsmath,amssymb]{revtex4}

\usepackage{graphicx}
\usepackage{dcolumn}
\usepackage{bm}

\begin{document}

\preprint{ArXiv/0611532}

\title{Cationic exchange in nanosized ZnFe$_2$O$_4$ spinel revealed by
experimental and simulated near-edge absorption structure}

\author{S.J. Stewart}
\email{stewart@fisica.unlp.edu.ar} \affiliation{IFLP-CONICET and
Departamento de F\'isica, Fac. Cs. Exactas, C.C. 67, Universidad
Nacional de La Plata, 1900 La Plata, Argentina}
\author{S.J.A. Figueroa}
\author{J. M. Ramallo-L\'opez}
\affiliation{Dto. F\'isica FCE. Universidad Nacional de La Plata and
IFLP-INIFTA (CONICET). 1900 La Plata.  Argentina.}
\author{S. G. Marchetti}
\author{J. F. Bengoa}
\affiliation{Departamento de Qu\'imica, Fac. Cs Exactas, Universidad
Nacional de La Plata, CONICET, CINDECA, CICPBA, 47 No 257, 1900 La
Plata, Argentina}
\author{R. J. Prado}
\affiliation{Departamento de F\'isica, ICET, UFMT, Av. Fernando
Corr\^{e}a s/n, CEP 78060-900, Cuiab\'a - MT, Brazil}
\author{F. G. Requejo}
\affiliation{Dto. F\'isica FCE. Universidad Nacional de La Plata and
IFLP-INIFTA (CONICET). 1900 La Plata.  Argentina.}
\date{\today}
\begin{abstract}
The non-equilibrium cation site occupancy in nanosized zinc ferrites
(6-13 nm) with different degree of inversion (0.2 to 0.4) was
investigated using Fe and Zn K-edge x-ray absorption spectroscopy
XANES and EXAFS, and magnetic measurements. The very good agreement
between experimental and ab-initio calculations on the Zn K-edge
XANES region clearly show the large
Zn$^{2+}$(A)$\rightarrow$Zn$^{2+}$[B] transference that takes place
in addition to the well-identified
Fe$^{3+}$[B]$\rightarrow$Fe$^{3+}$(A) one, without altering the
long-range structural order. XANES spectra features as a function of
the spinel inversion were shown to depend on the configuration of
the ligand shells surrounding the absorbing atom. This XANES
approach provides a direct way to sense cationic inversion in these
spinel compounds. We also demonstrated that a mechanical
crystallization takes place on nanocrystalline spinel that causes an
increase of both grain and magnetic sizes and, simultaneously,
generates a significant augment of the inversion.
\end{abstract}
\maketitle
\section{Introduction}
Several works have demonstrated that when normal spinel ferrite
ZnFe$_2$O$_4$ becomes nanosized displays a non-equilibrium cation
distribution amongst their tetrahedral (A) and octahedral [B] sites
that alters its long-range magnetic ordering and drastically
enhances its magnetic response \cite{1,2,3,4,5}. Iron (III)
occupancy of both A and B-sites in nanocrystalline ZnFe$_2$O$_4$ has
been extensively proved by M\"{o}ssbauer spectroscopy
\cite{1,2,6,7,8}, nuclear magnetic resonance \cite{5}, neutron
diffraction \cite{3}, x-ray absorption \cite{9,10,11,12}, and
indirectly throughout magnetic measurements \cite{2,3,4,13,14}. To a
much lesser extend, Extended X-ray Absorption Fine Structure (EXAFS)
studies at Zn K-edge suggest that Zn ions are transferred from its
equilibrium position (sites A) to B-sites when the particle size
decreases \cite{9,10,11}. Conversely, there is still some lack of
clarity concerning Zn non-equilibrium positions and its effects.
Some results indicates that Zn non-equilibrium location produces an
overpopulation of B-sites by both Zn and Fe ions \cite{10}, while
other claims that it distorts the spinel structure and brings about
an amorphous state \cite{9}. Thus, more studies are needed
concerning the less-explored Zn local geometry to completely
understand the origin of the non-equilibrium cation distribution in
nanosized ZnFe$_2$O$_4$. X-ray Absorption Near Edge Spectroscopy
(XANES) reflects the excitation process of a core electron to bound
and quasi-bound states near to the Fermi level. Even though its
intensity is around one order of magnitude larger than that of the
EXAFS region and involves a reduced photon energy range (50  eV), it
is dominated by multiple scattering interactions that make the
interpretation of XANES spectra substantially more complicated.
However, recently improved simulation tools based in ab-initio XANES
calculations such as code FEFF 8.2 \cite{15} can be employed to
better elucidate the XANES features and get some insight into the
compound electronic structure. Here we study the non-equilibrium
cation site occupancy in nanosized ZnFe$_2$O$_4$ and how it affects
the XANES spectra at both Fe and Zn K-edges. Throughout the very
good agreement between experimental and theoretical results on the
Zn K-edge region we show the potentialities of this technique to
straightforwardly sense the Zn$^{2+}$(A)$\rightarrow$Zn$^{2+}$[B]
transference, in addition to the well-determined
Fe$^{3+}$(A)$\rightarrow$Fe$^{3+}$[B] one that confer peculiar
magnetic properties to nanosized ZnFe$_2$O$_4$. We also show that
milling nanocrystalline ZnFe$_2$O$_4$ a considerable cation exchange
occurs that increases the degree of inversion as the grain and/or
particle sizes increase without altering the long-range structure.
\section{Experimental}
ZnFe$_2$O$_4$ nanoferrite(sample 2ZF) was prepared by precipitating
aqueous mixtures of zinc (II) and iron (III) nitrates in a ratio
Zn:Fe 1:2 with aqueous ammonia. The suspension was hydrothermally
processed in a Teflon-lined autoclave at 250$^\circ$C. The product
was filtered, washed with distilled water and dried. Samples 2ZF4h,
and 2ZF10h were obtained by high-energy ball milling nanocrystalline
2ZF for $t_m$= 4 and 10 hours, respectively, in a horizontal miller
(Retsch) with a stainless steel vial and one stainless steel ball
(mass to powder ratio was 10:1). Bulk ZnFe$_2$O$_4$ was prepared by
a conventional solid sate reaction. Transmission electronic
microscopy was performed in a JEOL JEM-200FX microscope operating at
200 keV. Magnetic measurements were carried out using both
commercial superconducting quantum interference device magnetometer
and AC susceptometer. $^{57}$Fe M\"{o}ssbauer spectra in the 4.2 to
300 K range of temperature were taken in transmission geometry.
EXAFS and XANES spectra of the Fe K edge (7112 eV) and Zn K edge
(9659 eV) were recorded at room temperature in transmission mode
using a Si(111) monochromator with a slit aperture of 0.3 mm at the
XAS beamline of the LNLS (Laboratório Nacional de Luz S\'incrotron)
in Campinas, Brazil.
\section{Results and discussion}
Previously \cite{12}, X-ray diffraction studies showed that all of
the samples are of single phase ZnFe$_2$O$_4$ cubic spinel. No
amorphous components were detected. In addition, we have observed
that milling causes a progressive increment of the average grain
size from 6 to 13 nm \cite{12}. Dark field TEM micrographs showed
particles in the 4-20 nm range, with average sizes of 6 nm and 8 nm
for 2ZF and 2ZF10h, respectively. Atomic absorption spectrometry
showed that the initial Zn:Fe ratio of 1:2 was unaltered by milling.
M\"{o}ssbauer spectra recorded at 4.2 K showed that Fe$^{3+}$ ions
occupy both A and B magnetic sublattices \cite{12}. Further, we
observed that the relative area ratio site(A):site[B] doubles from
2ZF to 2ZF10h, indicating a milling-induced transference of iron
ions from B to A-sites. The thermal evolution of the M\"{o}ssbauer
spectra (not shown) associated with susceptibility measurements
(Fig. 1) provided evidence for a superparamagnetic relaxation of
particle moments. The blocking temperature, T$_B$, taken as the
maximum of the in-phase ac susceptibility, $\chi$', is 38, 135 and
220 K for 2ZF, 2ZF4h and 2ZF10h, respectively (Fig. 1). The
progressive increment of T$_B$ reflects the magnetic size growth
with the milling. The S-shapes displayed by the M-H curves (Fig. 1)
evidence magnetic states with spontaneous magnetization. Further,
the magnetic response rather increases by increasing the milling
time as seen by the increment of the saturation magnetization M$_S$
(at T=5 K, M$_S$= 35, 57 and 68 emu/g for 2ZF, 2ZF4h and 2ZF10h,
respectively). There is also an increment of the magnetic hardness
with the grain size (the coercive fields are 87 and 198 Oe for 2ZF
and 2ZF10h, respectively).
\begin{figure}[-h]
\includegraphics[width=8cm]{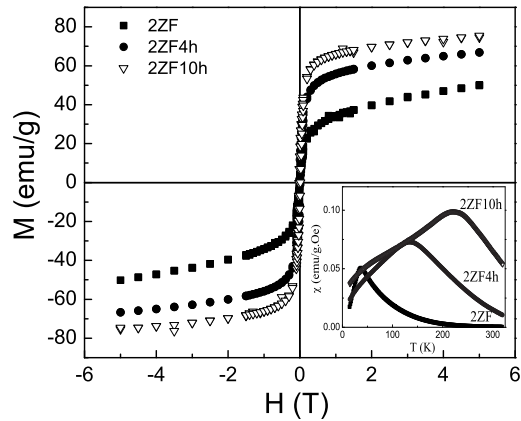}
\caption{Magnetization versus applied magnetic field (M-H) loops
taken at 5 K under a maximum field of 5 T for 2ZF, 2ZF4h and 2ZF10h
nanosized zinc ferrites. Inset: thermal dependence of the in-phase
ac susceptibility.}
\end{figure}
These magnetic and M\"{o}ssbauer results are compatible with a
ferrimagnetic state of the nanoferrites.\\The $\chi$(k) EXAFS signal
was extracted using the Athena program and analyzed using the
Arthemis program \cite{16}. Fourier transform (FT) of $\chi$(k) at
Zn and Fe K-edges without phase correction are shown in Fig. 2 and
fit results are summarized in Tables I and II, respectively.
\begin{figure}[-h]
\includegraphics[width=5cm]{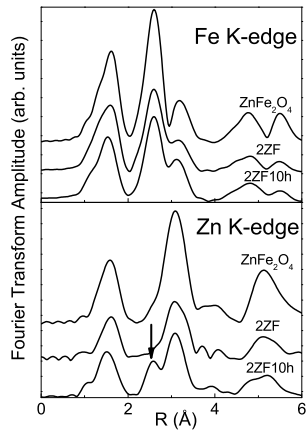}
\caption{Fourier transforms of the EXAFS spectra Fe and Zn K-edges
of bulk ZnFe$_2$O$_4$  and nanosized ferrites 2ZF and 2ZF10h. Arrow
indicates the new coordination sphere because large Zn population at
B sites.}
\end{figure}
Similarly to previous works \cite{9,10,11}, there is a reduction of
the amplitude of the main peaks when compared to bulk ZnFe$_2$O$_4$,
associated with the local structural disorder at particle surface
layer. Ideally, in normal ZnFe$_2$O$_4$ the zinc atoms are
tetra-coordinated by oxygens at 1.996 {\AA} and 12 Fe atoms are at
3.500 {\AA} as second neighbors. On the other side, iron atoms are
surrounded by 6 O at 2.018 {\AA} and 6 Fe atoms at 2.995 {\AA} as
second neighbors. We observe that the EXAFS spectra resemble that of
the bulk material evidencing that the local structure is almost
preserved in nanosized ZnFe$_2$O$_4$ (Tables I and II).
\begin{table}[-ht]
\caption{\label{table1}EXAFS Fe K-edge fitted parameters. $N$ is the
coordination number, $R$ the distance from central atom, $\sigma$
the Debye-Waller factor, $E_{0}$ the energy shift and $c$ is the
degree of inversion determined according to the Fe K-edge EXAFS
fitted data. Errors are indicated in parenthesis.}
\begin{ruledtabular}
\begin{tabular}{ccccccc}
Sample&Shell&N&R(\AA)&$\sigma^2$[\AA$^{-2}$]${\times}10^{-2}$&$E_{0}(eV)$&$c$\\
\cline{1-7}\\
2ZF&O&5.5(9)&1.96(3)&0.7(2)&-5(2)&22\\
2ZF10h&O&5.8(9)&1.97(3)&0.8(2)&-5(2)&38\\
\end{tabular}
\end{ruledtabular}
\end{table}
\begin{table}[-ht]
\caption{\label{table2}EXAFS Zn K-edge fitted parameters.\\{*}A
third cumulant was used to fit these shells. Values obtained were
7(2)${\times}10^{-4}$ and 8(3)${\times}10^{-4}$ for 2ZF and 2ZF10h ,
respectively.}
\begin{ruledtabular}
\begin{tabular}{ccccccc}
Sample&Shell&N&R(\AA)&$\sigma^2$[\AA$^{-2}$]${\times}10^{-2}$&$E_{0}(eV)$&$c$\\
\cline{1-7}\\
&O*&4.6(8)&2.01(2)&0.66(2)&6(1)&20\\
2ZF&Fe&13(2)&3.51(1)&1.08(2)&6(1)&-\\
&O&13(2)&3.54(1)&1.08(2)&6(1)&-\\
&Fe&4.4(8)&3.66(1)&1.08(2)&6(1)&-\\
\\
\cline{1-7}\\
&O*&4.9(8)&2.04(2)&0.9(2)&3.6(8)&44\\
&Fe&0.8(4)&2.94(3)&0.3(1)&-2.6(8)&-\\
2ZF10h&Fe&11(2)&3.49(1)&1.0(2)&3.6(8)&-\\
&O&11(2)&3.52(1)&1.0(2)&3.6(8)&-\\
&Zn&3.8(8)&3.62(1)&1.0(2)&3.6(8)&-\\
\end{tabular}
\end{ruledtabular}
\end{table}
However, an additional peak appears in the FT of the Zn K-edge
spectrum of 2ZF10h sample (Fig. 2, see arrow). When Zn atoms occupy
B-sites, a new shell of Fe atoms would appear at R=2.980 {\AA} from
central Zn. Thus, the similar coordinate R obtained from our fit
(Table II) indicates that the new peak corresponds to Zn
octahedrally coordinated, as reported in References. \cite{9,10,11}.
The fact that this peak is clearly resolved indicates a large cation
transference Zn$^{+2}$(A)$\rightarrow$Zn$^{+2}$[B] in 2ZF10h. From
these EXAFS analysis at both Fe and Zn K-edges we have determined
the degree of inversion, $c$, that represents the site
occupancy as (Zn$_{1-c}$Fe$_c$)[Zn$_c$Fe$_{2-c}$]O$_4$, where ( ) and\\
{[ ]} refer to A and B sites, respectively (Tables I and II).
Further, the values obtained from Zn and Fe edges independently are
both in good agreement with each other. This discards any site
overpopulation in our samples. All the above results show that
milling nanosized ZnFe$_2$O$_4$ produces an increment of both the
grain and magnetic sizes but keeping a nanometric characteristic
length, and, at the same time, the inversion increases. We
demonstrate that the augment of the degree of inversion in
ZnFe$_2$O$_4$ is not directly associated with a surface effect due
to the grain size reduction but with the synthesis process,
something that was not clearly stated in previous studies
\cite{2,3,4,11}. In particular, high-energy ball-milling process
activates an order-disorder transformation independently whether it
is causing an increment or a reduction of the grain/particle sizes.
In the following, we analyze how the near-edge x-ray absorption
structure at Fe and Zn K-edges are affected by the spinel inversion.
Figures 3 and 4 show the Fe-K and Zn-K XANES spectra for nanosized
and bulk ZnFe$_2$O$_4$ respectively. The peak broadening is probably
related to cation distribution amongst locally disordered sites
whose energy transition varies according to the average bond length
\cite{17}.
\begin{figure}[-h]
\includegraphics[width=5cm]{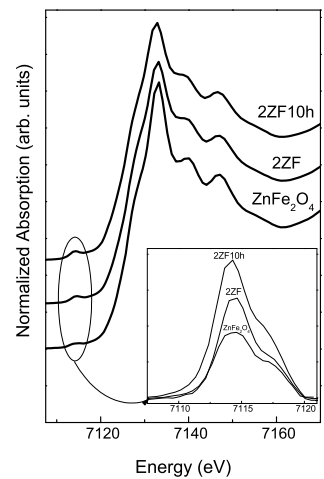}
\caption{Experimental XANES spectra at the Fe K-edge of 2ZF, 2ZF10h
and bulk ZnFe$_2$O$_4$. The inset shows the pre-edge region, after
discounting the edge jump contribution.}
\end{figure}
Fe-K pre-edge structure arises from electronic 1s$\rightarrow$3d
quadrupole and 1s$\rightarrow$3d/4p (hybridized orbitals) dipole
transitions \cite{18}. In agreement with M\"{o}ssbauer results, the
pre-edge position is compatible with a Fe$^{3+}$ oxidation state
\cite{19}.The integrated intensity of the pre-peaks region
progressively increases for 2ZF, 2ZF4h and 2ZF10h with respect to
the bulk compound (see inset in Fig. 3). This is due to the
increment of the degree of orbital p-d mixing as more central Fe
atoms occupy non-centrosymmetric A-sites as follows:
2ZF${<}$2ZF4h${<}$2ZF10h. Above the edge, the decrease in amplitude
of the white line would also indicate the presence of Fe$^{3+}$ in a
four-coordinated surrounding \cite{17}, although, no other
remarkable differences were detected in the Fe-K XANES region.
\begin{figure}[-h]
\includegraphics[width=5cm]{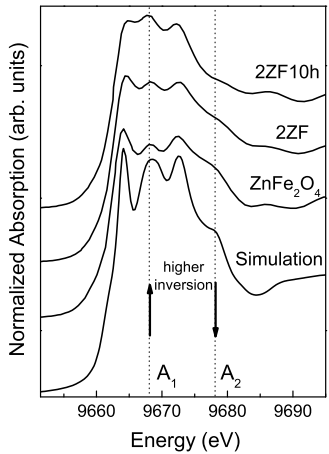}
\caption{Experimental XANES spectra at the Zn K-edge of nanosized
2ZF and 2ZF10h, bulk ZnFe$_2$O$_4$ and simulation result of the
XANES spectrum of normal ZnFe$_2$O$_4$ calculated using FEFF 8.2.}
\end{figure}
The most striking features due to cation exchange in ZnFe$_2$O$_4$
appear at the Zn-K edge (Fig. 4), which reflects the electronic
transition from zinc 1s core level to unoccupied states of p type.
Zn-K XANES of normal ZnFe$_2$O$_4$ consists of three resolved peaks
at 9664, 9666.3 (see line A$_1$ in Figs. 4 and 5) and 9672.4 eV, and
a significant shoulder at 9676.7 eV (see line A$_2$ in Figs. 3 and
4) at the white line region plus additional structure at higher
energies (Fig. 4) \cite{20}. The spectrum of 2ZF nanoferrite shows
some differences at the second feature and the shoulder region (Fig.
4). Furthermore, by increasing inversion the peak positions slightly
shift to lower energies, the second feature notably increases in
amplitude and the shoulder weakens. To further study the inversion
effects on the Zn-K electronic structure, we have performed XANES
simulations using FEFF 8.2 code \cite{15}. Here, we have considered
the Hedin-Lundqvist exchange potential with imaginary part of 0.5 eV
to simulate experimental broadening \cite{21}. The Z + 1
approximation (Z, atomic number) for the absorber atom was essential
to account for the three main peaks and thus reproduce accurately
the experimental data. A 57 atoms cluster (radius of 0.57 nm) was
used to calculate the self-consistent field (SCF) muffin-tin atomic
potential, and a 185 atoms cluster (radius of 0.805 nm) for the
full-multiple scattering (FMS) XANES calculations. For convergence
analysis, simulations considering up to 390 atoms (radius of 1.000
nm) and 112 atoms (radius of 0.650 nm) for the FMS and SCF
calculations were performed, respectively, and they did not show
differences from the results shown here. During the simulation
process, the normal spinel ZnFe$_2$O$_4$ was used as model
structure. The simulated absorption spectrum of the Zn atoms in
tetrahedral site was obtained using this normal structure. On the
other way, to calculate the spectrum of the Zn absorber in
octahedral site, a Zn atom substituted a Fe atom of the model
structure, and this Zn atom was used as the absorber during
calculations. Figure 4 shows the theoretical XANES spectrum of
normal ZnFe$_2$O$_4$. The theoretical data was shifted in order
obtain a better agreement between the experimental and theoretical
ZnFe$_2$O$_4$ spectra, and all the other theoretical data shown here
were shifted by this same value. The very good agreement between
experimental and simulated data is evident just by inspection on the
main features of the absorption spectra. In addition, the same
tendency with the degree of inversion (i.e. transferring Zn atoms
from sites with $T_d$ to $O_h$ symmetry) can be observed in
simulated set of XANES spectra. Indeed, Fig. 5 shows the theoretical
calculation for an isolated Zn atom substituting Fe at an octahedral
site in normal ZnFe$_2$O$_4$ structure (Fig. 5 (a)).
\begin{figure}[-h]
\includegraphics[width=5cm]{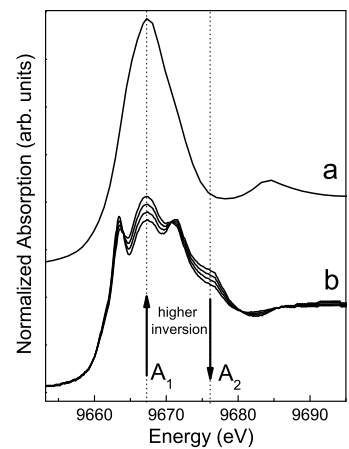}
\caption{Theoretical calculation for isolated Zn atom
substitutionally replacing Fe at B sites in ZnFe$_2$O$_4$ structure
(a) and theoretical ZnFe$_2$O$_4$ spectra using weights of 0.0, 0.1,
0.2 and 0.3 for Zn at B sites. (b).}
\end{figure}
It is worth noticing the coincidence between the positions of the
second peak of the ZnFe$_2$O$_4$ structure (line $A_1$) and the
white line of the Zn substitutional to Fe in ZnFe$_2$O$_4$
structure. The substitution of the second and/or fourth coordination
shells around the absorber atom, by Zn or Fe atoms, give no
significant changes, but causes slight variations in the relative
heights of the three main peaks near the edge. The effect on
theoretical XANES spectra of increasing Zn population at B sites in
ZnFe$_2$O$_4$ (Fig. 5(b)) was calculated by the weighted sum of both
normal spinel spectrum (Zn in tetrahedral site, shown in Fig. 4) and
the spectrum of a Zn absorber substitutional to Fe atom, at the
octahedral B sites (Fig. 5 (a)). The four spectra in figure 5 (b)
were obtained using weights of 0.0, 0.1, 0.2 and 0.3 for the Zn in
octahedral B sites (the degree of inversion), and consequently
weights of 1.0, 0.9, 0.8 and 0.7 for Zn in normal tetrahedral A
sites. Arrows in Figure 5 indicate the tendency of each feature with
increasing inversion. The results show not only the intensity
enhancement at 9666.3 eV (see line A$_1$ in Figs. 4 and 5) but also
the amplitude reduction of the shoulder near 9676.7 eV (see line
A$_2$ in Figs. 4 and 5), with increasing concentration of Zn at the
otherwise Fe equilibrium location in ZnFe$_2$O$_4$. Both trends
coincide with those observed experimentally, showing the important
local structural changes in nanostructured zinc ferrite that involve
a considerable zinc cation transference from A to B sites that
modifies its electronic structure. It is worth mentioning that when
we consider a weight of 0.5 for the Zn in octahedral B sites the
main features of the XANES disappears and the white line dominates
the spectrum shape.
\section{Conclusions}
Summarizing, chemically prepared ZnFe$_2$O$_4$ nanocrystalline
displays an inverted structure that changes its long-range magnetic
order and enhances its magnetic response with respect to bulk
antiferromagnetic material. This is in accordance with all previous
results on nanosized ZnFe$_2$O$_4$ independently on the synthesis
procedure. We demonstrate that the mechanical crystallization that
takes place on nanocrystalline spinel, which involves both grain and
apparent magnetic size growth, is also accompanied by a significant
augment of the inversion. This result shows that the inversion
involves the whole particle and is not only restricted to a more
distorted surface layer. On the other hand, XANES results give a
direct proof of the non-equilibrium cation distribution through the
Fe K-edge pre-edge features but mainly through Zn K-edge features,
which change due to the particular configuration of the ligand
shells surrounding the absorbing atom. These features are
intensified by the cationic swap originated by transferences of type
Zn$^{2+}$(A)$\rightarrow$Zn$^{2+}$[B] and
Fe$^{3+}$[B]$\rightarrow$Fe$^{3+}$(A) that take place preserving the
structural long-range order of the compound. Thus, superexchange
interactions Fe$^{3+}$[B]${-}$O$^{-2}$${-}$Fe$^{3+}$(A) emerge and
confer ZnFe$_2$O$_4$ a cluster glass or ferrimagnetic magnetic
behavior \cite{3}. The large zinc occupancy of octahedral sites that
we found cause broken superexchange A-B paths by the presence of
non-magnetic ions (Zn$^{2+}$) that, in addition to an inhomogenously
distributed inversion, bring about regions where ferrimagnetic A-B
or antiferromagnetic B-B interactions coexist \cite{5}. On the other
hand, considering that the magnetic properties of ferrimagnetic
material are critically dependent on site occupancy, XANES approach
as outlined here promises to be an important tool in determining the
structure of these materials. Further, our results encourage future
effort to understand the electronic structure throughout the unusual
Zn K-edge near-edge features of ZnFe$_2$O$_4$ \cite{20} or
investigate suggested electron density anomalies in nanosized
ZnFe$_2$O$_4$ \cite{22}.
\\
\begin{acknowledgments}
We appreciate financial support by LNLS synchrotron, Campinas - SP,
Brazil (project D04B - XAFS1 4148/05); ANPCyT, Argentina (PICT03
06-17492); CONICET, Argentina (PIP 6524 and PIP 6075). DC magnetic
measurements were performed using the RN3M facilities. We thank E.
D. Cabanillas for TEM microscope operation and F. Sives for fruitful
discussions.
\end{acknowledgments}


\begin{thebibliography}{99}
\bibitem{1}H. H. Hamdeh, J. C. Ho, S. A. Oliver, R. J. Willey, G. Oliveri, G. Busca, J. Appl. Phys. 81, 1851 (1997)
\bibitem{2}C. N. Chinnasamy, A. Narayanasamy, N. Ponpandian, K. Chattopadhyay, H. Guérault, J-M. Greneche, J. Phys. Condens. Matter 12, 1 (2000)
\bibitem{3}M. Hofmann, S. J. Campbell, H. Ehrhardt, R. Feyerherm, J. Mater. Sci. 39, 5057 (2004)
\bibitem{4}M. K. Roy, B. Haldar, H. C. Verma, Nanotechnology 17, 232 (2006)
\bibitem{5}J. H. Shim, S. Lee, J. H. Park, S-J Han, Y. H. Jeong, Y. W. Cho, Phys. Rev. B 73 064404 (2006)
\bibitem{6}F. S. Li, L. Wang, Q. G. Zhou, X. Z. Zhou, H. P. Kunkel, G. Williams, J. Magn. Magn. Mat. 268, 332 (2004)
\bibitem{7}S. A. Oliver, H. H. Hamdeh, J. C. Ho, Phys. Rev. B 60, 3400 (1999)
\bibitem{8}E. J. Choi, Y. Ahn, K. C. Song, J. Magn. Magn. Mat., in press
\bibitem{9}B. Jeyadevan, K. Tohji, K. Nakatsuka, J. Appl. Phys. 76, 6325 (1994)
\bibitem{10}S. A. Oliver, V. G. Harris, H. H. Hamdeh, J. C. Ho, Appl. Phys. Lett. 76, 2761 (2000)
\bibitem{11}S. Ammar, N. Jouini, F. Fiévet, O. Stephan, C. Marh  ic, M. Richard, F. Villain, Ch. Cartier dit Moulin, S. Brice, Ph. Sainctavit, J. Non-Cryst. Solids, 345-346, 658 (2004)
\bibitem{12}S. J. Stewart, S. J. A. Figueroa, M. B. Sturla, R. B. Scorzelli, F. Garc\'ia, F. G. Requejo, Physica B, doi:10.1016/j.physb.2006.07.045 (2006)
\bibitem{13}A. Kundu, C. Upadhyay, H. C. Verma, Phys. Lett. A 311, 410 (2003)
\bibitem{14}R. D. K. Misra, S. Gubbala, A. Kale, W. F. Egelhoff Jr., Mat. Scie. Lett. B 111, 164 (2004)
\bibitem{15}J. J. Rehr, R.C. Albers, C.R. Natoli, E.A. Stern, Phys. Rev. B 34, 4350 (1986)
\bibitem{16}B. Ravel, M. Newville J. Synchrotron Rad. 12, 537 (2005)
\bibitem{17}G. A. Waychunas, M. J. Apted, G. E. Brown Jr,  Phys. Chem. Minerals 10, 1-9 (1983)
\bibitem{18}F. de Groot, Chem. Rev. 101, 1779 (2001)
\bibitem{19}M. Wilke, F. Farges, P-E. Petit, G. E. Brown Jr., F. Martin, American Mineralogist 86, 714 (2001)
\bibitem{20}G. A. Waychunas, C. C. Fuller, J. A. Davis, J. J. Rehr, Geochim. Cosmochi. Acta 67, 1031 (2003)
\bibitem{21}M. Roy and S. J. Gurman, J. Synchrotron Rad. 8, 1095 (2001)
\bibitem{22}C. Upadhyay, H. C. Verma, Appl. Phys. Lett. 85, 2074 (2004)
\end{thebibliography}
\end{document}